\documentclass[12pt]{article}
\usepackage{epsfig}
\usepackage{graphicx}
\usepackage{lineno}
\usepackage{rotating}
\usepackage{lscape}
\usepackage{amsmath}
\usepackage[usenames]{color}
\usepackage{color}

\begin{document}
\begin{center}
\vspace*{1.0cm}

{\Large \bf{The half-life of $^{212}$Po}}

\vskip 1.0cm

{\bf P.~Belli$^{a,b}$, R.~Bernabei$^{a,b,}$\footnote{Corresponding
author. {\it e-mail address:} rita.bernabei@roma2.infn.it
(R.~Bernabei).}, R.S.~Boiko$^{c,d}$, F.~Cappella$^{e,f}$,
V.~Caracciolo$^{a,b,g}$, R.~Cerulli$^{a,b}$, F.A.~Danevich$^{c}$,
A.~Incicchitti$^{e,f}$, D.V.~Kasperovych$^{c}$,
V.V.~Kobychev$^{c}$, O.G.~Polischuk$^{c}$, N.V.~Sokur$^{c}$,
V.I.~Tretyak$^{c}$}

\vskip 0.3cm

$^{a}${\it INFN sezione Roma ``Tor Vergata'', I-00133 Rome, Italy}

$^{b}${\it Dipartimento di Fisica, Universit$\grave{a}$ di Roma
``Tor Vergata'', I-00133 Rome, Italy}

$^{c}${\it Institute for Nuclear Research of NASU, 03028 Kyiv,
Ukraine}

$^{d}${\it National University of Life and Environmental Sciences
of Ukraine, 03041 Kyiv, Ukraine}

$^{e}${\it INFN sezione Roma, I-00185 Rome, Italy}

$^{f}${\it Dipartimento di Fisica, Universit$\grave{a}$ di Roma
``La Sapienza'', I-00185 Rome, Italy}

$^{g}${\it INFN, Laboratori Nazionali del Gran Sasso, I-67100
Assergi (AQ), Italy}

\end{center}

\vskip 0.5cm
\begin{abstract}
The half-life of $^{212}$Po was measured with the highest
up-to-date accuracy as $T_{1/2}=295.1(4)$ ns by using
thorium-loaded liquid scintillator.
\end{abstract}

\vskip 0.2cm


\section{Introduction}

\begin{figure}[!ht]
 \begin{center}
 \mbox{\epsfig{figure=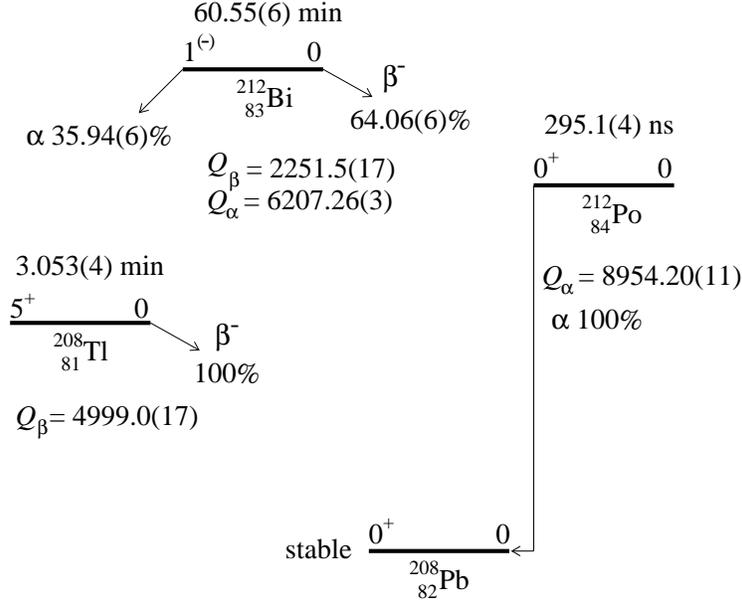,height=8.0cm}}
\caption{The decay chain
$^{212}$Bi$\rightarrow$$^{212}$Po$\rightarrow$$^{208}$Pb
\cite{NDS-212}. The $\alpha$ transition of $^{212}$Bi to
$^{208}$Tl is shown too \cite{NDS-208}. The half-life of
$^{212}$Po is from the present study. Energies of $\beta$ and
$\alpha$ decays are given in keV.}
 \label{fig:decay-scheme}
 \end{center}
 \end{figure}

The $^{212}$Po nuclide is the $\alpha$ active daughter of the
$^{212}$Bi from the $^{232}$Th decay chain with the shortest decay
time among the naturally occurring radioactive nuclides. The decay
scheme of the
$^{212}$Bi$\rightarrow$$^{212}$Po$\rightarrow$$^{208}$Pb (BiPo)
and $^{212}$Bi$\rightarrow$$^{208}$Tl sub-chains is shown in
Fig.~\ref{fig:decay-scheme}. The current recommended value of the
$^{212}$Po half-life is $T_{1/2}=294.3(8)$ ns \cite{NDS-212}. The
history of the $^{212}$Po half-life measurements is presented in
Table \ref{tab:half-life}.

\begin{table*}[!ht]
\caption{Historical view of the measured half-lives of
$^{212}$Po.}
\begin{center}
\small
\begin{tabular}{|l|l|l|}
 \hline
 Reference                      & Experimental technique / Compilation                  & Half-life, ns \\
 (year)                         &                                                       &  \\
 \hline
 \cite{Dunworth:1939} (1939)    & Geiger-M\"{u}ller counters, external source           & $300(100)$   \\
 \hline
 \cite{Bradt:1943} (1943)       & Geiger-M\"{u}ller counters, external source           & $260(40)$ \\
 \hline
 \cite{Hill:1948} (1948)        & Geiger-M\"{u}ller counters, external source           & $300(15)$   \\
 \hline
 \cite{Bunyan:1949} (1949)      & Geiger-M\"{u}ller and proportional counters,          & \\
 ~                              & external source                                       & $304(4)$   \\
 \hline
 \cite{VanName:1949} (1949)     & Geiger-M\"{u}ller counters, external source           & $220(10)$ \\
 \hline
 \cite{Hayashi:1953} (1953)     & Geiger-M\"{u}ller and proportional counters,            &  \\
 ~                              & external source                                       & $290(10)$ \\
 \hline
 \cite{Flack:1962} (1962)       & Plastic scintillators, external source                & $305(25)$   \\
 \hline
 \cite{Astner:1963} (1963)      & CsI(Tl) and plastic scintillators, external source    & $305(5)$ \\
 \hline
 \cite{McBeth:1972} (1972)      & Source in liquid scintillator                         & $302(6)$$^1$   \\
 \hline
 \cite{Sanyal:1975} (1975)      & Plastic scintillator, surface barrier Au-Si detector, & \\
                                & external source                                       & $296(2)$   \\
 \hline
 \cite{Bohn:1981} (1981)        & Surface barrier Au-Si and HP-planar Ge detectors,     &   \\
   ~                            & external source                                       & $309(11)$   \\
 \hline
 \cite{Bellini:2013} (2013)     & Source in liquid scintillator                         & $294.7(10)$ \\
 \hline
 \cite{Belli:2014} (2014)       & Source in BaF$_2$ scintillator                        & $298.8(16)$ \\
 \hline
 \cite{Aprile:2017} (2017)      & Source in liquid/gas Xe time projection chamber       & $293.9(12)$ \\
 \hline
 \cite{Belli:2018} (2018)       & Source in liquid scintillator                         & ~ \\
 ~                              & (1st stage of the current experiment)                 & $294.8(19)$ \\
 \hline
 \cite{NDS-212} (2020)          & Nuclear Data Sheets compilation                       & $294.3(8)$ \\
 \hline
 Present study                  & Source in liquid scintillator                         & $295.1(4)$ \\
 \hline

 \multicolumn{3}{l}{$^1$ Average of the two values $300(8)$ ns and $304(8)$ ns reported in Ref. \cite{McBeth:1972}.}\\

\end{tabular}
\normalsize
\end{center}
\label{tab:half-life}
\end{table*}

Gaseous counters have been used in the early experiments
\cite{Dunworth:1939,Bradt:1943,Hill:1948,Bunyan:1949,VanName:1949,Hayashi:1953}
to detect the $\beta$ particle emitted in the $^{212}$Bi decay and
the subsequent $\alpha$ particle of $^{212}$Po. Combinations of
scintillation and semiconductor detectors were utilized in the
experiments \cite{Flack:1962,Astner:1963,Sanyal:1975,Bohn:1981},
with $\beta$ and $\alpha$ particles from an external source (where
the BiPo sequence of decays occurred) registered by
the detectors.

A different approach has been exploited by using a liquid
scintillator enriched with a source containing the BiPo chain  \cite{McBeth:1972}.
A similar approach was exploited by the Borexino collaboration:
quartz vials with thorium and $^{220}$Rn-loaded liquid
scintillators were inserted into the Borexino Counting Test
Facility (CTF) detector \cite{Bellini:2013}. The half-life of
$^{212}$Po was also measured in the experiment \cite{Belli:2014}
with the help of a BaF$_2$ crystal scintillator contaminated by
radium. However, the accuracy of the experiment was limited by the
comparatively slow scintillation response of the BaF$_2$
scintillator (effective scintillation decay time
$\tau_{eff}\approx0.6$ $\mu$s) and a poor signal-to-noise ratio
due to a rather modest light yield of the scintillator ($\simeq
10^3$ photons/MeV$_{\gamma}$) \cite{BaF2}. In the XENON100
detector (a xenon liquid/gas time projection chamber) a $^{220}$Rn
source was used to calibrate it; using such calibration data, a
further measurement of the $^{212}$Po half-life was obtained
\cite{Aprile:2017}.

It should also be noted that the half-life of $^{212}$Po was
measured in Refs.
\cite{Belli:2003,2003Da24,Cerulli:2004,Belli:2007,2012Be14,2018Sa45,2018So16}
as a by-product of the BiPo sub-chain analysis for different
purposes (estimation of low-counting detectors' internal
contamination by thorium, calibration of detectors, study of
low-lying states in $\alpha$-decaying nuclei). Typically the
results are affected by large statistical uncertainties up to
17\%\footnote{It should be noted, however, a rather small
uncertainty of the half-life value obtained in work
\cite{Cerulli:2004}: $T_{1/2}=297(1)$ ns.}, and the systematic
effects were not evaluated. Much larger statistics was gathered in
recent measurement \cite{Alexeev:2018} where an external Th source
was placed between plastic scintillators viewed by a
photomultiplier (PMT) obtaining $T_{1/2}=294.09(7)$ ns.
Unfortunately, systematic effects in the experiment were not
estimated. In Ref. \cite{Ma:2020} a radon gaseous source was used
to calibrate the xenon liquid/gas time-projection chamber of the
PandaX-II experiment obtaining $T_{1/2}=297(6)$ ns with quite big
uncertainty comparing with those in the last experiments in Table
\ref{tab:half-life}.

Considering the data in Table 1, the liquid scintillator loaded
with Th radionuclides appears a very promising way to reduce the
$^{212}$Po half-life uncertainty. However, although the
measurements performed by the Borexino collaboration have profited
of a fast scintillation signal, the measurement's precision was
limited by the following aspects: the time jitter of the 100
photomultipliers array, scattering of the
scintillation photons, their absorption and re-emission inside
the CTF volume, and the readout electronics bandwidth. Moreover, a
substantial difference ($\sim40$ cm) in the photon paths to reach
the PMTs has induced an additional time-spread distribution
(average value of ~2 ns) due to the large volume of the CTF
detector (4.8 m$^3$ liquid scintillator).

According to all these observations, in order to improve the
accuracy on the $^{212}$Po half-life measurements, one should
minimize the liquid scintillator volume and use a fast response
PMT/electronics. A thorium-loaded liquid scintillator (LS(Th)) was
developed for the present study. Preliminary results of the
measurements were published in \cite{Belli:2018}. Here we report
the final result of the experiment with $\approx64$ times larger
statistics.

\section{Experiment}
\label{sec:exp}

\subsection{Thorium-loaded liquid scintillator}
\label{sec:LS-Th}

\subsubsection{Production of thorium-loaded liquid scintillator}
\label{sec:ls-development}

Thorium nitrate pentahydrate Th(NO$_3)_4\cdot5$H$_2$O was used as
initial Th compound to prepare the LS(Th). A 20\% solution of
trioctylphosphine oxide (TOPO) in toluene was taken as complexing
organo\-phosphorous agent to bind thorium in organic phase. The
mixture was stirred with thorium nitrate pentahydrate salt to
obtain a solution containing 2 mg of Th in 1 mL of TOPO solution:

 \begin{center}
 Th$^{4+}~+~4$NO$^{-}_{3}$~+~3$\overline{\mathrm{TOPO}} \rightleftharpoons
 \overline{\mathrm{Th(NO}_{3})_{4}(\mathrm{TOPO})_{3}}$.
 \end{center}

The obtained Th-containing organic solution was diluted 20-fold
with the liquid scintillator based on toluene with the addition of
0.1\% 2.5-diphenyl oxazole (PPO) and 0.01\%
1,4-bis(5-phenyloxazol-2-yl) benzene (POPOP). Taking into account
the preparation procedure, the liquid scintillator contains
$\approx0.1$ wt\% of thorium ($^{232}$Th and $^{228}$Th with
daughters), while the activity of $^{228}$Ra is expected to be
rather low. Moreover, while $^{228}$Ra has too small energy of
$\beta$ decay ($Q_{\beta}=45.5(6)$ keV \cite{Wang:2020}) well
below the hardware threshold (see Sec. \ref{sec:deltat}), presence
of some amount of $^{228}$Ac ($\beta$ active daughter of
$^{228}$Ra with the decay energy $Q_{\beta}=2123.8(26)$ keV
\cite{Wang:2020}) leads to a mild contribution to the random-pairs
background. Thus presence of some amount of $^{228}$Ra in the
scintillator almost does not affect the half-life of $^{212}$Po
determination.

\subsubsection{Scintillation properties and activity of $^{232}$Th and its daughters in the LS(Th)}
\label{sec:ls-prop}

A 7-mL sample of the LS(Th) was sealed inside a quartz vial with
inner sizes $(\oslash33\times14)$ mm to test its scintillation
properties and the activities of $^{232}$Th and $^{228}$Th with
daughters. The vial was optically connected to a PMT Philips
XP2412 and covered by 3M reflector foil to improve the
scintillation-light collection. The signals from the PMT after a
preamplification stage entered a shaping amplifier with 0.5 $\mu$s
shaping time and then were processed by a peak sensitive
analog-to-digital converter.

The relative light yield of the scintillator was estimated by
using $^{137}$Cs and $^{207}$Bi $\gamma$ sources to be 42(3)\% in
comparison to a polystyrene based plastic scintillator
$\oslash30\times15$ mm measured in the same conditions.

\begin{figure}[!htbp]
 \begin{center}
 \mbox{\epsfig{figure=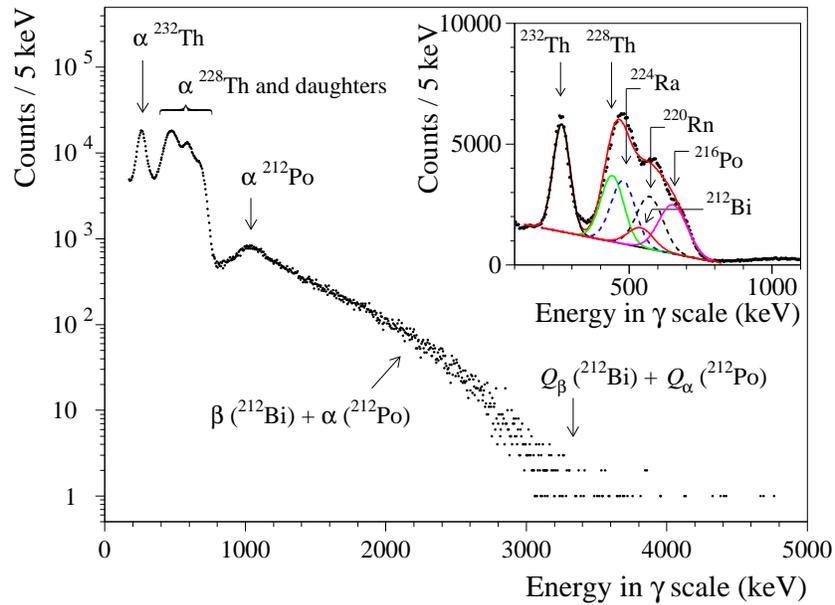,height=8.0cm}}
 \caption{Energy spectrum taken with 7 mL LS(Th)
scintillator over 5910~s. The main features of the spectrum are
shown. (Inset) Fit of the energy spectrum accumulated with the
LS(Th) over 2050 s by a model which includes $\alpha$ peaks of
$^{232}$Th and $^{228}$Th with daughters.}
 \label{fig:ls-bg}
 \end{center}
 \end{figure}

The energy spectrum measured with the LS(Th) is presented in Fig.
\ref{fig:ls-bg}. The energy scale of the detector was determined
by analysis of the Compton electron spectra of the $^{137}$Cs and
$^{207}$Bi $\gamma$ sources as suggested in \cite{Dietze:1982}.
The peaks, in the (0.25--0.7) MeV energy range (see Fig.
\ref{fig:ls-bg}), can be attributed to $\alpha$ decays of
$^{232}$Th and $^{228}$Th with daughters. The broad peak at
$\sim1.1$ MeV is due to the $\alpha$ decays of $^{212}$Po when the
energy of the previously emitted $\beta$ particle is low. The
overlap of $\beta$ events of $^{212}$Bi and $\alpha$ events of
$^{212}$Po produces the broad energy distribution up to $\sim3.3$
MeV because the used acquisition did not separate these two events
occurred in too short time in comparison to the amplifier shaping
time.

\nopagebreak

To estimate the activity of $^{232}$Th and $^{228}$Th with
daughters, the energy spectrum accumulated with the LS(Th) has
been fitted by a model built from the $\alpha$ peaks of
$^{232}$Th, $^{228}$Th, $^{224}$Ra, $^{220}$Rn, $^{216}$Po and
$^{212}$Bi. The activities of the radionuclides, the energy
resolution and the $\alpha/\gamma$ ratio\footnote{Defined as
position of $\alpha$ peak in the energy scale measured with
$\gamma$ quanta. Here we neglect the contribution of nuclear
recoils because of a much higher quenching \cite{Tretyak:2014}.}
were free parameters of the fit. A linear function has been used
to describe the distribution of $\beta$ particles and $\gamma$-ray
quanta. Furthermore, taking into account the radon escape
possibility from the LS(Th), we introduce a coefficient to
consider a lower activity of $^{220}$Rn and  its daughters. The
result of the fit in the energy interval $165-785$ keV
($\chi^2$/n.d.f. = 1.45, where n.d.f. is the number of degrees of
freedom) is presented in the Inset of Fig. \ref{fig:ls-bg}. The
obtained behavior of the $\alpha/\gamma$ ratio in the energy range
(4--9) MeV is described by the following formula:
$\alpha/\gamma=0.02149(14)+0.01104(3)\times E_{\alpha}$ (where
$E_{\alpha}$ is energy of the $\alpha$ particles in MeV). The
activities of $^{232}$Th and $^{228}$Th have been measured as
4.61(2) Bq/mL and 3.82(7) Bq/mL, respectively. The lower activity
of $^{228}$Th is due to its decay after the thorium compound
preparation in January 2016 (the measurements were performed on
July 8th, 2016). The activity of the $^{220}$Rn and its daughters
is $92(2)\%$ of the $^{228}$Th activity; this can be explained by
radon escape from the LS(Th). The properties of the LS(Th) are
summarized in Table \ref{table:prop}.

 \begin{table}[!ht]
\caption{Properties of the thorium-loaded liquid scintillator.}
\begin{center}
\begin{tabular}{|l|l|l|}
\hline
 Property                                       & Value     & Note \\
 \hline
 Light yield                                    & 42(3)\%   & Relatively to polystyrene based  \\
 ~                                              & ~         & plastic scintillator \\
 Activity of $^{232}$Th & 4.61(2) Bq/mL         & ~ \\
 Activity of $^{228}$Th & 3.82(7) Bq/mL         & Reference date July 8th, 2016 \\
 Total $\alpha$ activity & 20.7(10) Bq/mL       & Reference date July 8th, 2016 \\
 Concentration of thorium & 0.113(1) wt\%    & ~ \\
 \hline
\end{tabular}
\end{center}
\label{table:prop}
\end{table}

 \subsection{Recording of BiPo waveforms}
 \label{sec:msr}

A sample of the LS(Th) in the same quartz vial was viewed by a
fast-time-response PMT Hamamatsu R13089-100-11 with rise time 2.0
ns, transit time 20 ns and transit time spread 170 ps (full width
at half maximum, FWHM). The quartz vial was covered by
polytetrafluoroethylene tape to improve the light collection. The
signals waveforms were recorded by a LeCroy WavePro 735Zi-A
oscilloscope with a sampling frequency 20 GSa/s and a 3.5 GHz
bandwidth.

The experiment was performed in two stages. In the first stage,
carried out in June 2017, a 4.4 g (5 mL) sample of the liquid
scintillator was utilized (the results of the first stage are
published in \cite{Belli:2018}). A 10.6 g (12 mL) sample of the
LS(Th) was used in the second stage of the experiment in April-May
2018. In total 785548 events were recorded in the first stage,
while 50340611 events (larger by a factor of $\approx64$) were
recorded in the second stage of the experiment during 216.67 hours
in order to improve the half-life value precision. The
oscilloscope energy threshold in the second stage was set above
the $^{232}$Th $\alpha$ peak in order to reduce the counting rate.

\section{Data analysis and results}
\label{sec:da}

\subsection{Time intervals between $\beta$ and $\alpha$ pulses in BiPo events}
\label{sec:deltat}

\begin{figure}[!ht]
 \begin{center}
 \mbox{\epsfig{figure=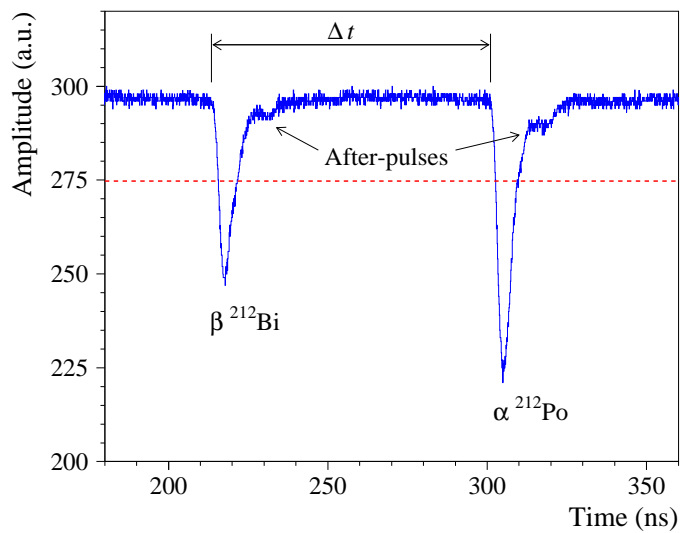,height=7.0cm}}
\caption{Example of $\beta$ pulse of $^{212}$Bi and subsequent
$\alpha$ pulse of $^{212}$Po in the liquid scintillator loaded by
thorium. Dashed red line shows the software threshold used to
identify pairs of events in the data. The time interval between
the signals is denoted as $\Delta t$.}
 \label{fig:bi-po}
 \end{center}
 \end{figure}

An example of a pair event in the LS(Th) classified as a BiPo
event is shown in Fig. \ref{fig:bi-po}. After-pulses both for
$\beta$ and $\alpha$ pulses of approximately $12-14$ ns after the
main pulses arise in the PMT because of the elastic scattering of
the accelerated photoelectrons on the 1st dynode. The scattered
electrons return to the photocathode and then are multiplied again
\cite{Hamamatsu}.

\begin{figure}[!ht]
 \begin{center}
 \mbox{\epsfig{figure=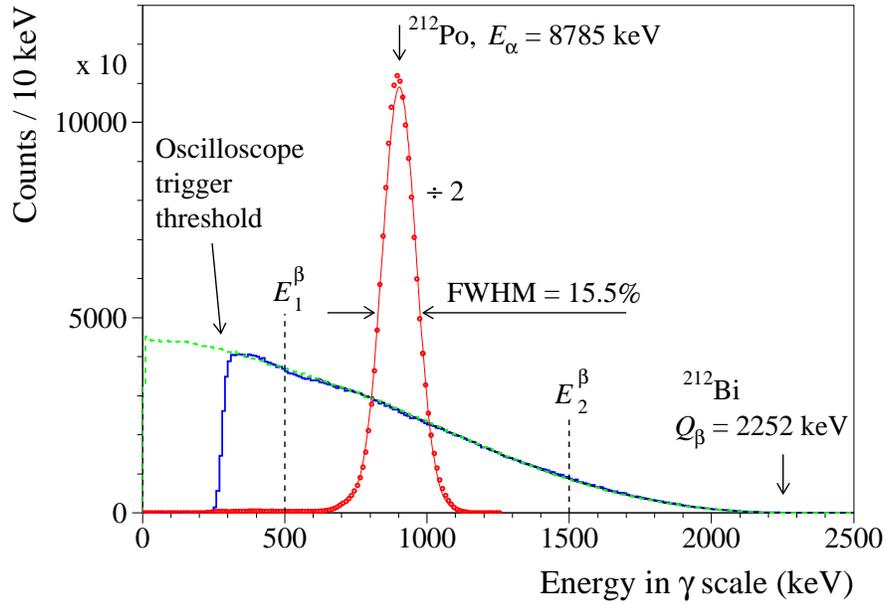,height=8.0cm}}
\caption{(Color online) Energy spectra of the 1st events ($\beta$
particles of $^{212}$Bi, solid histogram) and 2nd events ($\alpha$
particles of $^{212}$Po, dots). The Monte Carlo simulated response
function of the detector to the $\beta$ decay of $^{212}$Bi is
shown by dashed histogram. The energy interval of $\beta$ events
used in the analysis of the $^{212}$Po half-life is shown by
vertical dashed lines labelled $E^{\beta}_1$ and $E^{\beta}_2$.
The fit of the $\alpha$ peak by Gaussian function is shown by
solid line. The $\alpha$ peak is divided by a
factor of 2 to fit the figure.}
 \label{fig:b-a-sp}
 \end{center}
 \end{figure}

\begin{figure}[!ht]
 \begin{center}
 \mbox{\epsfig{figure=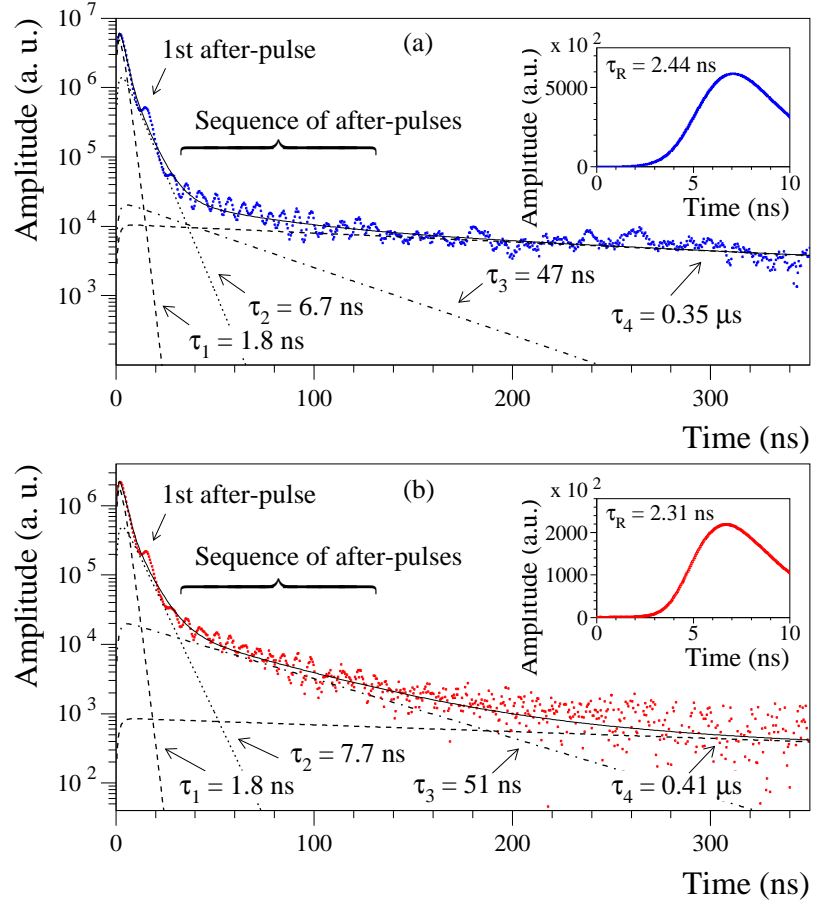,height=12.0cm}}
\caption{Shape of scintillation pulses in the LS(Th) for $\beta$
particles of $^{212}$Bi (a) and $\alpha$ particles of $^{212}$Po (b).
Fitting functions for $\beta$ and $\alpha$ pulses (solid lines),
with four components of the scintillation signals (dashed, dotted
and dash-dot lines) are shown. Zoomed $\beta$ and $\alpha$ pulses
are shown in Insets. $\tau_{R}$ denotes the rise time of the
pulses (see text for explanation).}
 \label{fig:pulse-shapes}
 \end{center}
 \end{figure}

Amplitude spectra of the first and second events are shown in Fig.
\ref{fig:b-a-sp}. The spectra were built by calculating the
signals area, after baseline subtraction, in a time interval 11 ns
(the time interval was chosen to avoid the effect of
after-pulses). A constant was taken to describe the baseline of
the first pulse, while a slow component of $\beta$ pulse (see Fig.
\ref{fig:pulse-shapes} and discussion of the scintillation signals
pulse shapes below) was added to the constant to reconstruct the
second pulse shape. The detector energy scale was determined by
comparison of the first events distribution with the Monte Carlo
simulated $\beta$ spectrum of $^{212}$Bi. The response of the
detector to the $\beta$ decay of $^{212}$Bi was simulated using
the {\sc Geant4} package version 10.4.p02 (Shielding PEN physics
list) \cite{Agostinelli:2003,Allison:2006,Allison:2016} with
initial kinematics given by the {\sc Decay0} event generator
\cite{DECAY0a,DECAY0b}. The simulated-distribution shape
reasonably agrees with the experimental data (see Fig.
\ref{fig:b-a-sp}). The energy resolution of the LS(Th) detector
was estimated for the $\alpha$ peak of $^{212}$Po as
FWHM$=15.5\%$\footnote{The difference in the $^{212}$Po
$\alpha$-peak position in the spectra presented in Figures
\ref{fig:ls-bg} and \ref{fig:b-a-sp} can be explained by different
methods to build the spectra: a shaping amplifier and a peak
sensitive analog-to-digital converter (the spectrum Fig.
\ref{fig:ls-bg}), and by using area of pulses for $< 11$ ns in the
recorded waveforms (Fig. \ref{fig:b-a-sp}).}.  The obtained energy
spectra confirm that the recognized pairs of events are mainly
BiPo events.

\clearpage

In addition to the after-pulses observed $12-14$ ns after the main
pulses, a rather long sequence of pulses is visible in a sum of a
big number of signals (see Fig. \ref{fig:pulse-shapes} where sums
of about ten thousands $\beta$ pulses of $^{212}$Bi and $\alpha$
pulses of $^{212}$Po are presented). Explanation of the multiple
after-pulses is problematic. However, we assume that this effect
is somehow related to the operation of the photomultiplier. While
the after-pulses have no impact on the determination of the 1st
signal origin ($\beta$-particle of $^{212}$Bi), they may affect
the determination of the start time of the 2nd signal
($\alpha$-particle of $^{212}$Po), and thus of the time interval
between the two signals, especially for small $\Delta t<(30-60)$
ns.

\clearpage

The pulse shapes of $\beta$ and $\alpha$ scintillation signals are
slightly different in the LS(Th). Fits of the pulses presented in
Fig. \ref{fig:pulse-shapes} were done by the function:

\begin{equation}
f(t)=\sum A_{i}(e^{-t/\tau _{i}}-e^{-t/\tau
_{0}})/(\tau_{i}-\tau_{0}),\qquad t>0,
\end{equation}

\noindent where $A_{i}$ are the relative intensities, $\tau_{i}$
are the decay constants for different light-emission components,
and $\tau_{0}$ is the time constant that describes the
scintillation signal rise and integration of the photomultiplier
and electronics ($\tau_{0}\approx1.6(2)$ ns both for $\beta$ and
$\alpha$ pulses). Four decay components were observed with
$\tau_{1}\approx 1.8$ ns, $\tau_{2}\approx 7$ ns, $\tau_{3}\approx
0.05$ $\mu$s and $\tau_{4}\approx 0.4$ $\mu$s with different
intensities for $\beta$ and $\alpha$ particles (see Table
\ref{table:scint-decay}). It should be noted that the present
investigation cannot be considered as analysis of the LS(Th)
pulse-shape, in particular since the after-pulses were not
excluded from the analysis. The estimations of the pulse-shape
decay profile were performed to take into account the slow
components of the $\beta$ pulses in the determination of the time
intervals between $\beta$ and $\alpha$ pulses in BiPo events.

\begin{table}[ht]
\caption{Decay time of the LS(Th) scintillator for $\beta$ and
$\alpha$ particles. The decay constants and their relative
intensities are denoted as $\tau_i$ and A$_i$, respectively.}
\begin{center}
\begin{tabular}{|l|l|l|l|l|}
\hline

 Type of events     &  \multicolumn{4}{|c|}{Decay constants (ns) and relative intensities}  \\

 \cline{2-5}

  ~                 & $\tau_1$, A$_1$  & $\tau_2$, A$_2$ & $\tau_3$, A$_3$ &  $\tau_4$, A$_4$ \\

\hline

 $\beta$ particles  & 1.8(4),            & 6.7(5),            & 47(3),             & 353(28),   \\

        ~           & $56(10)\%$        & $32(4)\%$         & $2.7(2)\%$        & $9.3(4)\%$ \\

 \hline

 $\alpha$ particles & 1.8(4),            & 7.7(6),            & 51(3),             & 409(134),   \\

        ~           & $52(9)\%$         & $37(7)\%$         & $7.7(4)\%$        & $3.3(2)\%$ \\

\hline

\end{tabular}
\end{center}
\label{table:scint-decay}
\end{table}

The following method was developed to find pairs of events in the
data and determine the time intervals $\Delta t$ between the
pulses:

1) A simple low-level-discriminator algorithm with a high
threshold (see Fig. \ref{fig:bi-po}) was applied to find pairs of
events in the data.

\begin{figure}[!ht]
 \begin{center}
 \mbox{\epsfig{figure=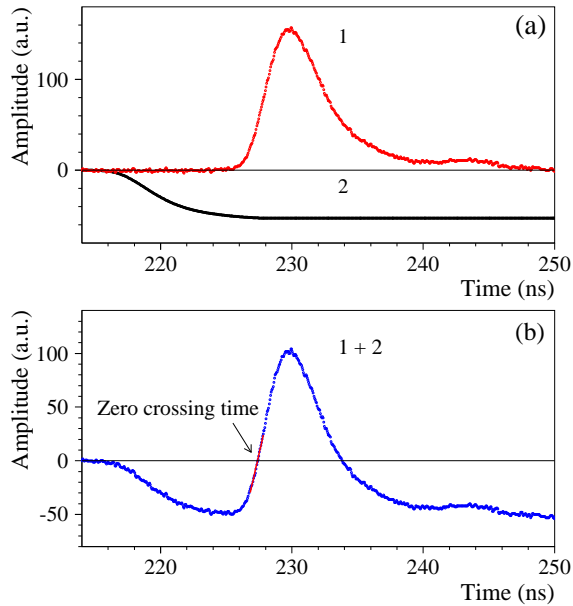,height=8.0cm}}
 \caption{Illustration of the digital constant-fraction discrimination
method used to determine the signals starting time and to
calculate $\Delta t$. (a) A scintillation pulse (after the
baseline subtraction) inverted and shifted by 11 ns (1); the pulse
integrated and multiplied by a factor 0.003 (2). (b) Sum of the
pulses 1 and 2. The solid line shows the fit of the data by an
exponential function. The arrow shows the zero crossing time
accepted as the signal starting time.}
 \label{fig:CFD}
 \end{center}
 \end{figure}

2) The recognized pairs of events were then analyzed by using the
method of digital constant-fraction discrimination illustrated in
Fig. \ref{fig:CFD}. In the method the pulse-origin time (denoted
in Fig. \ref{fig:CFD} as ``Zero crossing time'') was determined by
analyzing the sum of two pulses produced from the recorded signal
after the baseline subtraction: the first pulse was inverted and
shifted in time by 11 ns, and the second one integrated and
multiplied by a factor 0.003. The delay 11 ns was chosen to
minimize effect of the undelayed pulse fluctuations on the timing
(however, the delay was taken small enough to avoid possible
after-pulses effect on the integrated pulse). The zero crossing
time was found by fitting the sum pulse by exponential function
around zero value. An example of the fit is shown in Fig.
\ref{fig:CFD} too. The approach allows to eliminate an amplitude
dependence of the pulse-time origin, which appears in the simple
low-level-discriminator algorithm\footnote{We have checked how the
obtained half-life value (see below Section \ref{sec:half-life})
depends on the constant-fraction discrimination method parameters.
The data production was performed by using a constant-fraction
discrimination method with the delay 1.8 ns and the multiplication
factor 0.4 for the undelayed pulse (without integration). A fit of
the obtained time distribution returns the half-life value
$T_{1/2} = 295.09(26)$ ns in agreement with the value $T_{1/2} =
295.10(26)$ ns obtained with the method described above.}.

The $\beta$ and $\alpha$ pulses have slightly different rise time
($\tau_R$, defined as a time interval of the $(10-90)\%$ rising
edge): $\tau_R=2.44(16)$ ns for $\beta$ particles and
$\tau_R=2.31(14)$ ns for $\alpha$ particles (see Insets in Fig.
\ref{fig:pulse-shapes})\footnote{Rise time of scintillation
detector depends on several factors: photodetector sensitivity and
time properties, the readout electronics bandwidth, scintillation
material, size and geometry of the scintillator and reflector,
energy and ionization density of particle (see, e.g.,
\cite{Papadopoulos:1997,Derenzo:2000,Derenzo:2014}). The
difference in the rise time for $\alpha$ and $\beta$ particles
observed in the present study can be explained by their different
energy distributions and ionization densities.}. The $\tau_R$
values were calculated for $\beta$ and $\alpha$ signals with
amplitudes in the energy interval $700-1050$ keV (in the energy
scale of $\beta$ particles). This difference produces a systematic
shift of $\Delta t$ that was estimated by using Monte Carlo
simulations. The experimental data on the rise and trailing time
distributions, and the energy spectra for $\beta$ and $\alpha$
events were taken as input parameters for the simulations. Noise
was superimposed on the generated pulses. The noise was generated
taking into account the baseline fluctuations of the recorded
waveforms. In total $10^4$ double pulses were generated with
$\Delta t = 100$ ns. Analysis of the generated data by the
constant-fraction-discrimination algorithm returned $\Delta t =
99.74$ ns. The difference 0.26 ns was added to the $\Delta t$
values for each BiPo event in the data production process.

Taking into account the difference of scintillation pulse shapes
for $\beta$ and $\alpha$ particles, the mean-time method was
applied to analyze the pulse profiles of the events. For each
signal, the numerical characteristic of its shape (mean time,
$\zeta$) was defined by using the following equation:

\begin{equation}
\zeta = \sum f(t_{k})  \times t_{k} / \sum f(t_{k}),
\end{equation}

\noindent where the sum is taken over the time channels $k$,
starting from the origin of signal up to 60 ns; $f(t_{k})$ is the
digitized amplitude (at the time $t_{k}$) of a given signal.

 \begin{figure}[!ht]
 \begin{center}
 \mbox{\epsfig{figure=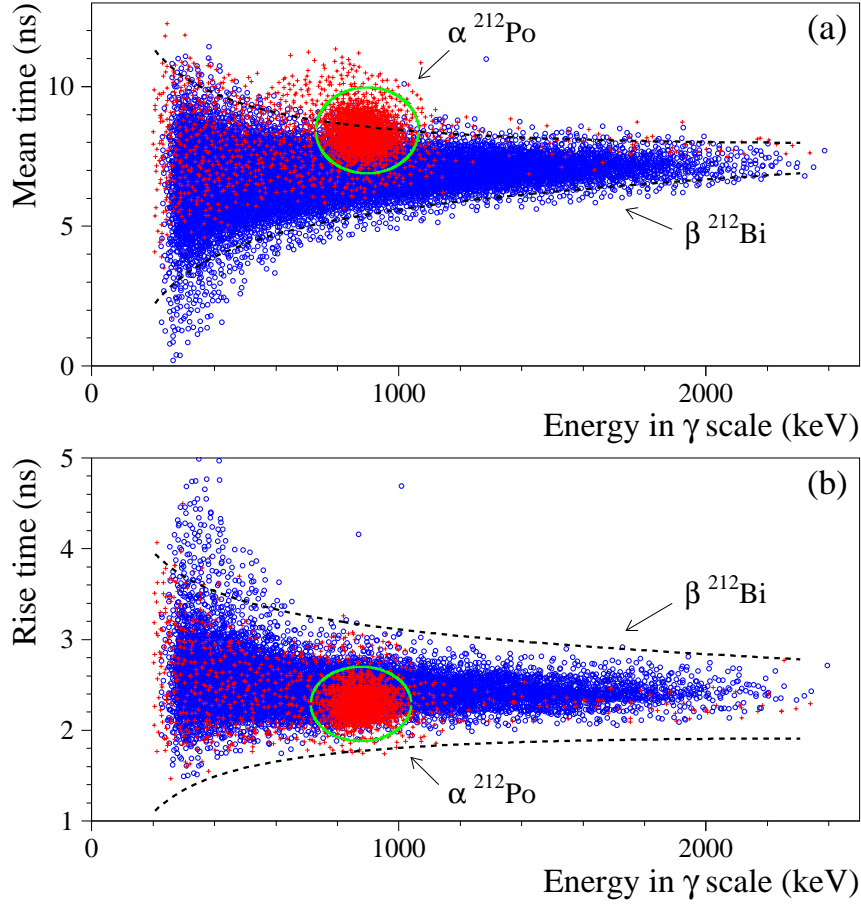,height=12.0cm}}
\caption{Mean time (a) and rise time (b) vs energy distributions
accumulated over 16.7 h with the LS(Th) (see text for explanation
of the parameters). The $\pm3\sigma$ intervals for the mean-time
and rise-time values for $\beta$ and $\alpha$ events selection are
depicted by dashed lines (for $\beta$ events) and by solid curves
in form of ellipses (for $\alpha$ particles).}
 \label{fig:MT-RT}
 \end{center}
 \end{figure}

The mean time vs energy distributions for the 1st ($\beta$
particles) and 2nd ($\alpha$ particles) events selected from the
data accumulated with the LS(Th) for 16.7 h are presented in Fig.
\ref{fig:MT-RT} (a). We have used $\pm3\sigma$ regions both for
$\beta$ particles of $^{212}$Bi and for the $\alpha$ peak of
$^{212}$Po to select the BiPo events for the further analysis. The
intervals to select $\beta$ and $\alpha$ events are shown in Fig.
\ref{fig:MT-RT}. Despite a rather poor particle discrimination
ability, the filter reduces the contribution of random pairs of
events and discards irregular-shape pulses of different origin.

 \begin{figure}[!ht]
 \begin{center}
 \mbox{\epsfig{figure=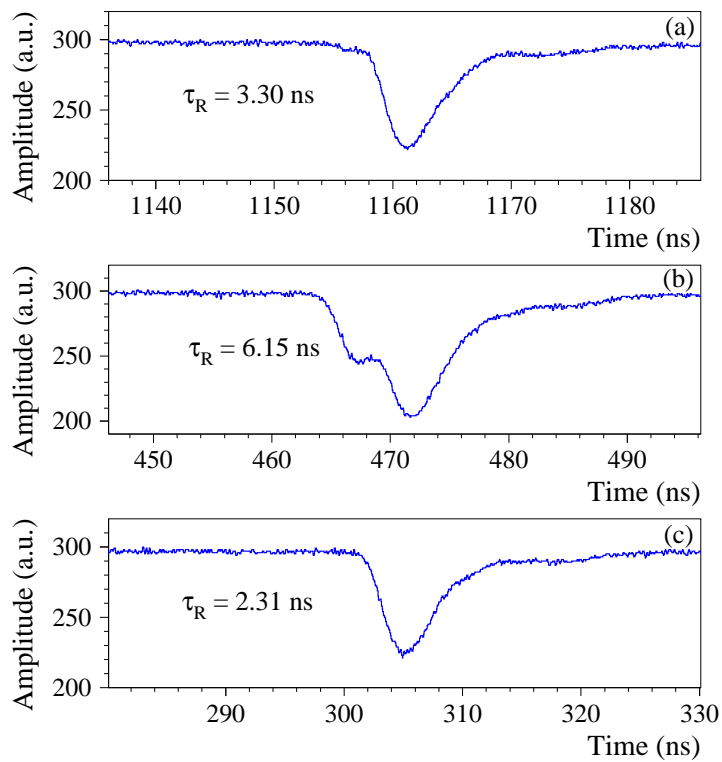,height=10.0cm}}
\caption{Examples of $\alpha$ pulses with the rise-time values
beyond the region of the accepted values for $\alpha$ particles
((a) and (b)). A pulse with a normal rise time also shown (c).}
 \label{fig:abnormal-RT}
 \end{center}
 \end{figure}

The rise-time parameters for $\beta$ and $\alpha$ events were also
analyzed (see Fig. \ref{fig:MT-RT} (b)). All the pairs with
$\beta$ or $\alpha$ pulses having the rise time outside the
$\pm3\sigma$ regions were discarded from the further analysis. Two
examples of $\alpha$ pulses with abnormal rise-time values are
shown in Fig. \ref{fig:abnormal-RT} (a) and (b) together with a
normal pulse (c). The pulses (a) and (b) with irregular shape can
be explained by overlap of scintillation pulses or overlap of
scintillation pulses with noise of different origin.

\subsection{Half-life of $^{212}$Po}
\label{sec:half-life}

\begin{figure}[!ht]
 \begin{center}
 \mbox{\epsfig{figure=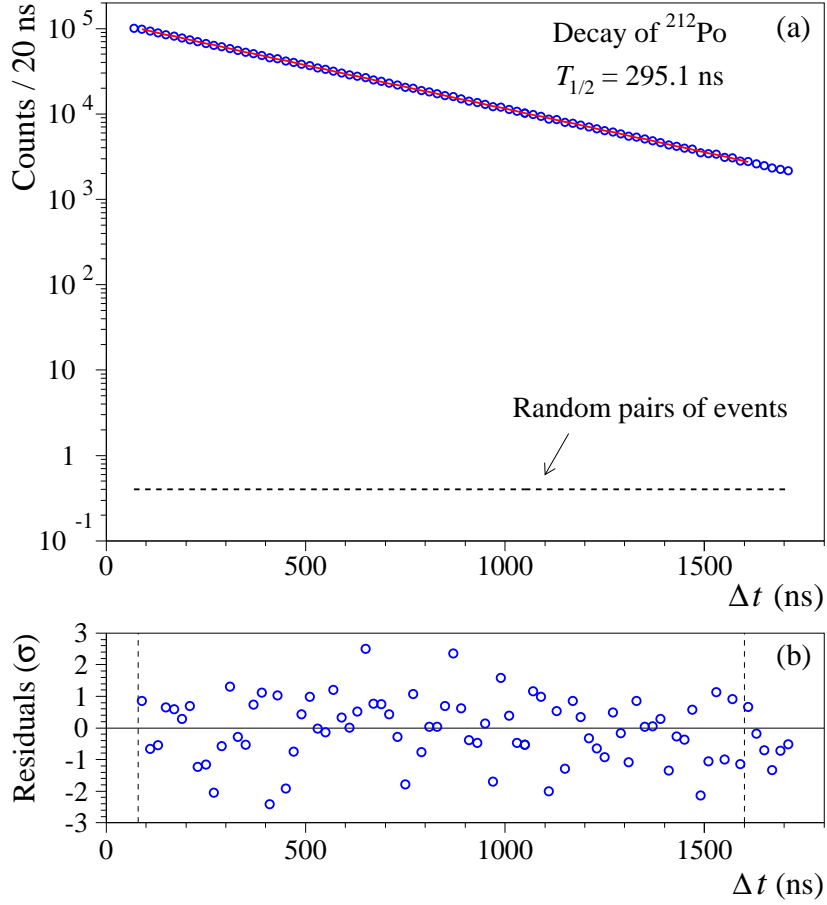,height=12.0cm}}
\caption{ (a) Distribution of the time intervals between the 1st
and 2nd signals in the selected pairs of events ($\Delta t$) and
its fit by an exponential function with the half-life
$T_{1/2}=295.1$ ns plus background due to random pairs of events.
The distribution of random pairs of events is shown too. (b)
Residuals of the fit. The interval of fit is shown by vertical
dashed lines.}
 \label{fig:time-int}
 \end{center}
 \end{figure}

The distribution of the time intervals between the first and
second signals in the recognized pairs of events is presented in
Fig. \ref{fig:time-int}. The pairs of events were selected with
the first events amplitudes inside the energy interval
$500-1500$~keV and with the mean-time and rise-time values inside
the $\pm3\sigma$ bands shown in Fig. \ref{fig:MT-RT}. The energy
interval to select the first events ($\beta$ particles of
$^{212}$Bi) for the analysis is also shown in Fig.
\ref{fig:b-a-sp} (the choice of the energy interval for the 1st
pulses selection will be explained below). The second events
($\alpha$ particles of $^{212}$Po) were accepted inside the
ellipses shown in Fig. \ref{fig:MT-RT}.

The time distribution was fitted by a sum of two exponential
functions: the first one describes the $\alpha$ decay of
$^{212}$Po and the second one is added to take into account random
pairs of events \cite{Radeloff:1967}:

\begin{equation}
N(t) = N_1~e^{(-t~\ln 2/T_{1/2})}+N_2~e^{(-t~b)},
 \label{eq:212po-decay}
 \end{equation}

\noindent where $N_1$ is proportional to the number of $^{212}$Po
$\alpha$ decays selected, $T_{1/2}$ is the half-life of
$^{212}$Po, $N_2$ is proportional to the number of random pairs of
events, and $b$ is an average rate of random pairs. The
random-pairs component (the parameters $N_2$ and $b$) was bounded
in the energy intervals chosen for the analysis from the counting
rate of the events taking into account the selection criteria for
the mean-time and rise-time parameters. It should be stressed that
the background due to the random pairs of events is very low in
the present experiment: the parameters $N_2$ and $b$ were
estimated to be $N_2\approx 3.3\times10^{-6}~N_1$ and
$b\approx2.4$ s$^{-1}$. The distribution of random pairs of events
is shown in Fig. \ref{fig:time-int} (a).

A fit of the $\Delta t$ distribution in the time interval
$80-1600$ ns by the maximum-likelihood method was performed with
the help of the PAW \cite{PAW} package that uses MINUIT software
\cite{MINUIT} for function minimization and uncertainty analysis.
The fit returns the half-life of $^{212}$Po 295.10(26) ns with
$\chi^2 = 77.4$ for 73 degrees of freedom. The results of the fit
are shown in Fig. \ref{fig:time-int}. According to the
recommendations \cite{GUM:2008}, taking into account that the
uncertainty of the half-life value was obtained by statistical
methods, it is standard uncertainty obtained by type A evaluation
method. Uncertainties due to possible systematic effects (Type B
evaluation of standard uncertainty \cite{GUM:2008}) and a combined
standard uncertainty of the result are reported in the next
Section.

\subsection{Combined standard uncertainty}

\begin{figure}[!ht]
 \begin{center}
 \mbox{\epsfig{figure=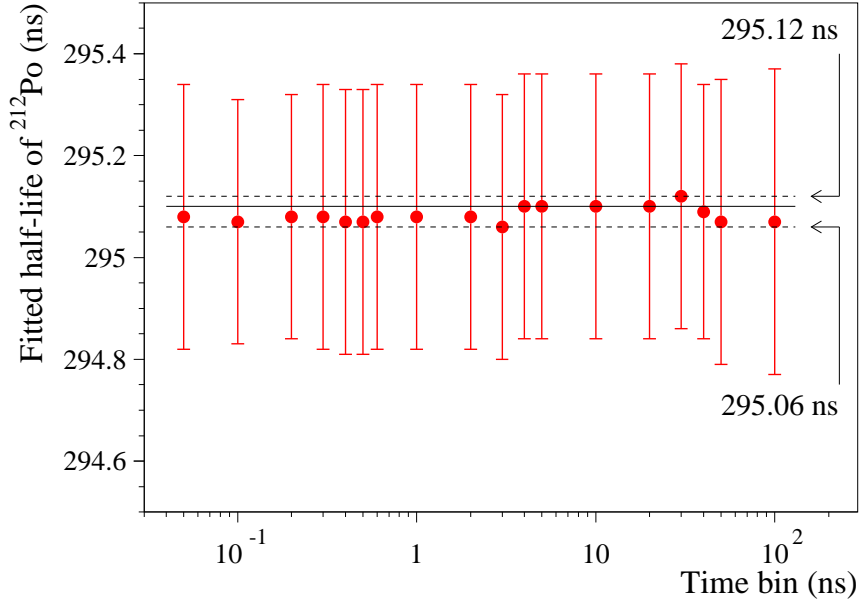,height=8.0cm}}
\caption{The half-life of $^{212}$Po as a function of the $\Delta
t$ distribution time-bin width. The value of the $^{212}$Po
half-life 295.1 ns is shown by solid line, while the upper and
lower values are shown by dashed lines.}
 \label{fig:bin-size}
 \end{center}
 \end{figure}

The $T_{1/2}$ values obtained by the fit of the data with the bin
widths from 0.05 ns (the oscilloscope time bin) to 100 ns lie
between 295.06 ns and 295.12 ns (see Fig. \ref{fig:bin-size}).
Assuming a uniform distribution of possible $T_{1/2}$ values
inside the interval, the standard deviation due to time-bin width,
0.02 ns, was calculated as the upper value
 minus the lower value divided by the
square root of 12 \cite{GUM:2008}.

\begin{figure}[!ht]
 \begin{center}
 \mbox{\epsfig{figure=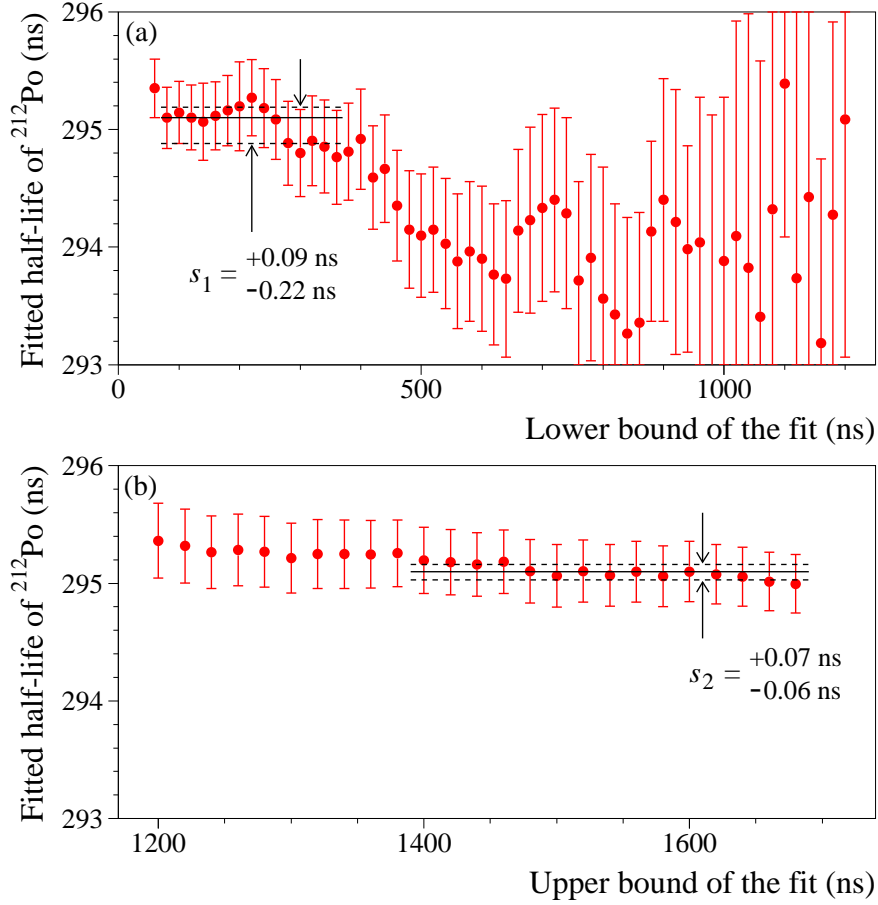,height=12.0cm}}
\caption{The half-life of $^{212}$Po as a function of the lower
(a) and higher bound (b) of the fit. 1600 ns was chosen as the end
point for the fits presented in the panel (a), while the starting
point was 80 ns for the fits shown in the panel (b). The standard
deviations (shown by dashed lines) were calculated for 15 values
in the time intervals 80--360 ns (denoted as $s_1$) and 1400--1680
ns ($s_2$) to estimate the uncertainties of the lower and upper
bounds of the fit. The value of the $^{212}$Po half-life
$T_{1/2}=295.1$ ns is shown by solid line.}
 \label{fig:int-of-fit}
 \end{center}
 \end{figure}

The result of fit depends on the low and high time bounds (see
Fig. \ref{fig:int-of-fit}). The interval of fit ($80-1600$ ns) was
chosen to minimize the combined uncertainty of the half-life
value, taking into account that a bigger time interval is
preferable to obtain smaller uncertainty due to statistical
fluctuations. However, the point 60 ns was excluded from the
analysis to reduce possible effect of after-pulses and slow
components of the scintillation decay on the determination of the
second pulse zero-crossing time\footnote{It should be noted that a
much bigger effect of the fitted half-life increase at low time
bounds was observed in the experiment \cite{Aprile:2017}.}.
Assuming a normal distribution of possible $T_{1/2}$ values, the
standard deviations of the 15 half-life values in the time
interval 80--360 ns were taken as uncertainties:
$s_1=_{-0.22}^{+0.09}$ ns (the 7 points above the value
$T_{1/2}=295.1$ ns were taken to calculate the upper uncertainty,
while the lower uncertainty was calculated by using the 7 points
below the value). Similarly the standard deviations of the 15 fit
results in the time interval 1400--1680 ns were taken as
uncertainties due to the upper bound of the fit interval:
$s_2=_{-0.06}^{+0.07}$ ns. The intervals where the uncertainties
were evaluated, as well as the lower and upper uncertainties, are
shown in Fig. \ref{fig:int-of-fit}.

\begin{figure}[!ht]
 \begin{center}
 \mbox{\epsfig{figure=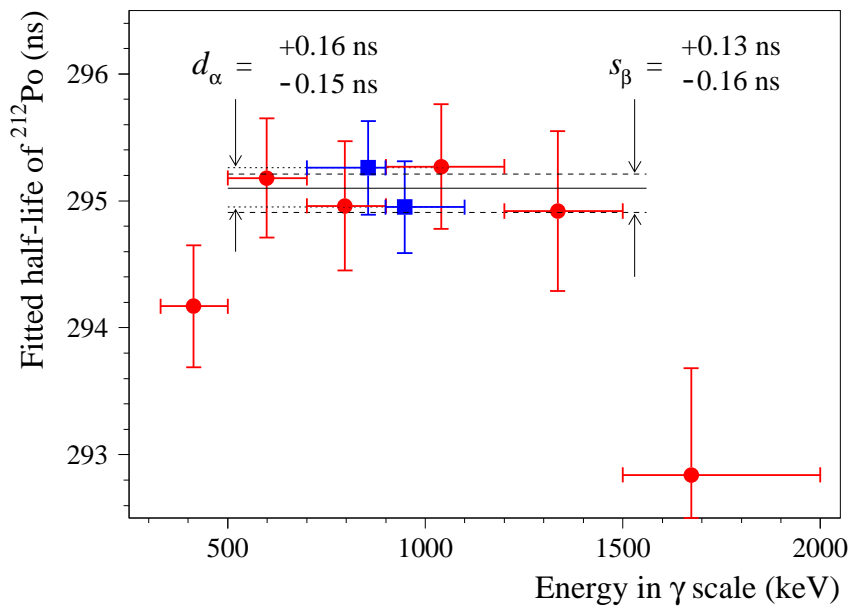,height=8.0cm}}
\caption{The half-life of $^{212}$Po as a function of the
amplitude of the 1st ($\beta$ particles of $^{212}$Bi, dots) and
2nd events ($\alpha$ particles of $^{212}$Po, squares). Dashed
lines show the uncertainties due to the $\beta$ pulses amplitude
(denoted $s_{\beta}$), dotted lines show the uncertainties due to
the $\alpha$ pulses amplitude ($d_{\alpha}$).}
 \label{fig:b-and-a-amplitude}
 \end{center}
 \end{figure}

There is an indication of the half-life dependence on amplitude of
the $\beta$ and $\alpha$ pulses. To estimate uncertainties due to
the $\beta$-events amplitude, the lower energy threshold of
$\beta$ events was varied within the energy interval 330--1500 keV
with the steps from 170 keV to 500 keV (see Fig.
\ref{fig:b-and-a-amplitude}). The dependence of the half-life on
the $\alpha$-signals amplitude was analyzed for the BiPo events
with $\alpha$ events in the first and second halves of the
$\alpha$ peak, while the $\beta$ events were in the energy
interval 500--1500 keV. The effect could be explained by
dependence of the PMT transit time on the pulse amplitude. The
related uncertainties were estimated as standard deviations within
the energy interval 500--1500 keV for $\beta$ particles
($s_{\beta}=_{-0.16}^{+0.13}$ ns), and as the differences
($d_{\alpha}=_{-0.15}^{+0.16}$ ns) between the half-life value
$T_{1/2}=295.1$ ns and the half-life values obtained for the two
halves of the $\alpha$ peak. Eventually the energy interval
500--1500 keV for the energy of $\beta$ particles was chosen for
the analysis to reduce a possible effect of the half-life
dependence on the pulses amplitude.

A possible effect of temperature variation during the data taking
was checked by analyzing the two data sets gathered in the time
intervals during the day with low (on the average 18$^{\circ}$ C
for 3.9 hours around 4:42 AM) and high (21$^{\circ}$ C for 3.9
hours around 1:36 PM) temperatures in the room where the
experimental set-up was installed. The analysis of the data sets
returned the half-life values $T_{1/2}=295.66(64)$ ns and
$T_{1/2}=295.43(62)$ ns for the ``low'' and ``high'' temperature
periods, respectively, that is no evidence for the effect.
Nevertheless, a systematic uncertainty due to a possible
temperature effects was estimated as the half of the difference
between the two values: $\pm0.12$ ns.

Finally, uncertainty of the oscilloscope to measure time interval
between two pulses was estimated as $\sqrt{2}\times SCJ + CAR$
\cite{LeCroy:2017}, where $SCJ$ is Sample Clock Jitter
($SCJ=0.0001$ ns) and $CAR$ is the product of the Clock Accuracy
(1 ppm) and the time stamp of the second pulse (conservatively we
took the maximal time interval 1600 ns and got the value
$CAR=0.0016$). Thus, we accept $0.0017$ ns as an uncertainty due
to a possible oscilloscope inaccuracy.

No other sources of uncertainty were observed. A summary of the
uncertainties is given in Table \ref{tab:syst}.

\begin{table}[!htbp]
\caption{Uncertainty evaluation of the half-life of $^{212}$Po
(ns).}
\begin{center}
\begin{tabular}{l|l}
 \hline
 Standard deviation by statistical methods  & $\pm0.26$ \\
 \hline
 Lower bound of the fit     & $_{-0.22}^{+0.09}$ \\
 \hline
 Upper bound of the fit     & $_{-0.06}^{+0.07}$ \\
 \hline
 Amplitude of $\beta$ events    & $_{-0.16}^{+0.13}$ \\
  \hline
 Amplitude of $\alpha$ events   & $_{-0.15}^{+0.16}$ \\
 \hline
 Time bin                       & $\pm0.02$ \\
 \hline
 Variations of temperature      & $\pm0.12$ \\
 \hline
 Uncertainty of the oscilloscope   & $\pm0.0017$ \\
 \hline
 Combined standard uncertainty  & $_{-0.43}^{+0.37}$ \\
 \hline
\end{tabular}
\end{center}
\label{tab:syst}
\end{table}
\normalsize

Treating all the listed in Table \ref{tab:syst} uncertainties as
independent and adding them in quadrature, we obtain the following
half-life of $^{212}$Po:

 \begin{center}
  $T_{1/2}=295.1(4)$ ns.
 \end{center}

\begin{figure}[!ht]
 \begin{center}
 \mbox{\epsfig{figure=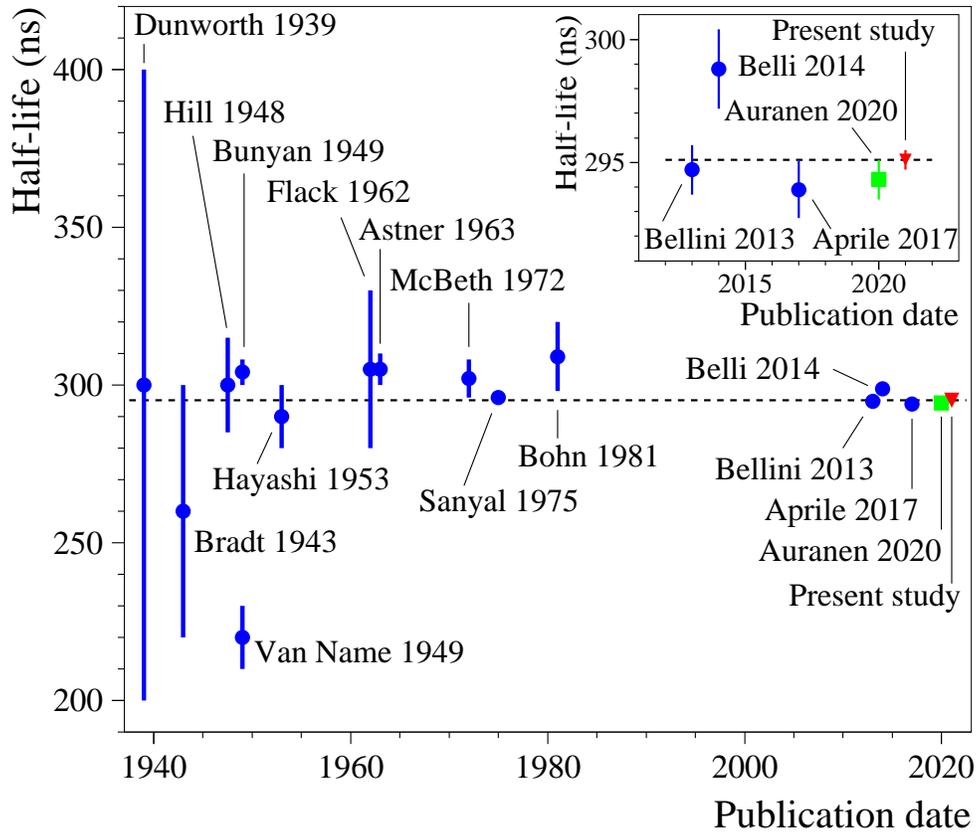,height=11.0cm}}
 \caption{A historical perspective of the half-life of
$^{212}$Po as a function of the publication date. References to
the publications are as follows: Dunworth 1939:
\cite{Dunworth:1939}, Bradt 1943: \cite{Bradt:1943}, Hill 1948:
\cite{Hill:1948}, Bunyan 1949: \cite{Bunyan:1949}, Van Name 1949:
\cite{VanName:1949}, Hayashi 1953: \cite{Hayashi:1953}, Flack
1962: \cite{Flack:1962}, Astner 1963: \cite{Astner:1963}, McBeth
1972: \cite{McBeth:1972}, Sanyal 1975: \cite{Sanyal:1975}, Bohn
1981: \cite{Bohn:1981}, Bellini 2013: \cite{Bellini:2013}, Belli
2014: \cite{Belli:2014}, Aprile 2017: \cite{Aprile:2017}. The
recommended value Auranen 2020 \cite{NDS-212} is shown by square;
the result of the present study is shown by triangle. In the Inset the most recent measurements
are reported.}
 \label{fig:history}
 \end{center}
 \end{figure}

The present measurement is the most accurate determination of the
$^{212}$Po half-life. The value is in agreement with the
preliminary result of the previous stage of the present experiment
$T_{1/2}=294.8(19)$ ns \cite{Belli:2018} and with the recommended
value $T_{1/2}=294.3(8)$ ns \cite{NDS-212}. The result agrees with
the recent experiments:
$T_{1/2}=(294.7\pm0.6(\mathrm{stat.})\pm0.8(\mathrm{syst.}))$ ns
\cite{Bellini:2013} and
$T_{1/2}=(293.9\pm1.0(\mathrm{stat.})\pm0.6(\mathrm{syst.}))$ ns
\cite{Aprile:2017}, but it is smaller than the one obtained in the
measurement with BaF$_2$ scintillation detector:
$T_{1/2}=(298.8\pm0.8(\mathrm{stat.})\pm1.4(\mathrm{syst.}))$ ns
\cite{Belli:2014}. A historical perspective of the half-life of
$^{212}$Po as a function of the publication date is presented in
Fig. \ref{fig:history}.

One can calculate a weighted average of the last half-life values
\cite{Bellini:2013}, \cite{Belli:2014}, \cite{Aprile:2017}, and of
the present study as $295.1(4)$ ns, with the uncertainties
combined in quadrature. If the value \cite{Belli:2014} is
excluded, the weighted average becomes $294.9(4)$ ns.

\section{Conclusions}

The half-life of $^{212}$Po relative to $\alpha$ decay to the
ground state of $^{208}$Pb (the only known channel of $^{212}$Po
decay) was measured with thorium-loaded liquid scintillator as
$T_{1/2}=295.1(4)$ ns. This result is the most accurate up-to-date
value (relative uncertainty: 0.14\%). It has been achieved thanks
to the utilization of the fast liquid scintillator (rise time does
not exceed $\sim 1$ ns), its small volume (12 mL), the use of the
fast photomultiplier with $\sim2$ ns rise time and of the fast
oscilloscope with a sampling frequency of 20 GSa/s and a 3.5 GHz
bandwidth, the high statistics of the acquired data, the developed
algorithm to determine with a high accuracy the time intervals
between the $\beta$ events of $^{212}$Bi and the $\alpha$ events
of $^{212}$Po.

\section{Acknowledgements}

The group from the Institute for Nuclear Research  of NASU (Kyiv,
Ukraine) was supported in part by the National Research Foundation
of Ukraine Grant No. 2020.02/0011. D.V.~Kasperovych,
O.G.~Polischuk and N.V.~Sokur were supported in part by the
project ``Investigations of rare nuclear processes'' of the
program of the National Academy of Sciences of Ukraine
``Laboratory of young scientists''.

\end{document}